# Large language models in finance : what is financial sentiment?

Kemal Kirtac[1] and Guido Germano[1,2]

9 March 2025

**Abstract:**
Financial sentiment has become a crucial yet complex concept in finance, increasingly used in market forecasting and investment strategies. Despite its growing importance, there remains a need to define and understand what financial sentiment truly represents and how it can be effectively measured. We explore the nature of financial sentiment and investigate how large language models (LLMs) contribute to its estimation. We trace the evolution of sentiment measurement in finance, from market-based and lexicon-based methods to advanced natural language processing techniques. The emergence of LLMs has significantly enhanced sentiment analysis, providing deeper contextual understanding and greater accuracy in extracting sentiment from financial text. We examine how BERT-based models, such as RoBERTa and FinBERT, are optimized for structured sentiment classification, while GPT-based models, including GPT-4, OPT, and LLaMA, excel in financial text generation and real-time sentiment interpretation. A comparative analysis of bidirectional and autoregressive transformer architectures highlights their respective roles in investor sentiment analysis, algorithmic trading, and financial decision-making. By exploring what financial sentiment is and how it is estimated within LLMs, we provide insights into the growing role of AI-driven sentiment analysis in finance.

## 1. Introduction

Financial sentiment refers to the collective attitude, emotions, and opinions expressed by investors, analysts, and the public toward financial markets. The concept of financial sentiment has been widely studied, particularly in the context of stock market prediction (Tetlock, 2007; Schumaker and Chen, 2009; Bollen et al., 2011; Loughran and McDonald, 2011; Heston and Sinha, 2017; Kirtac and Germano, 2024a,b).

[1] Department of Computer Science, University College London, 66–72 Gower Street, London WC1E 6EA, United Kingdom

Research has shown that sentiment measures extracted from textual sources, such as financial news, earnings reports, and social media, can be predictive of stock price movements (Kirtac and Germano, 2024a). Tetlock (2007) found that negative sentiment in financial news correlates with stock price declines, while Schumaker and Chen (2009) demonstrated that sentiment analysis of breaking news headlines can improve stock movement prediction. Other studies have explored how social media sentiment analysis impacts asset prices, with Twitter and StockTwits data often providing early signals of investor sentiment shifts (Bollen et al., 2011; Oliveira et al., 2017). More recently, Kirtac and Germano (2024a) examined the role of large language models (LLMs) in financial sentiment analysis and stock return prediction, finding that LLM-based approaches significantly outperformed traditional sentiment methods. Their sentiment-driven trading strategy demonstrated strong performance, highlighting the superior predictive power of LLMs compared to traditional lexicon-based sentiment models. Given the increasing availability of financial text data and the growing impact of machine learning in market prediction, the integration of sentiment indicators into algorithmic trading models has become a rapidly expanding area of research (Tetlock et al., 2008; Heston and Sinha, 2017).

The increasing reliance on quantitative sentiment measures has led to a rapid rise in studies using textual analysis and machine learning to predict financial market trends (Tetlock et al., 2008; Heston and Sinha, 2017). Sentiment analysis is now commonly integrated into algorithmic trading models, where real-time news feeds and investor sentiment indices inform buy and sell decisions (Rao and Srivastava, 2014; Karoui, 2017). Given this trend, there has been a large influx of financial sentiment-related papers, with a focus on leveraging deep learning and LLMs to enhance accuracy in market predictions (Loughran and McDonald, 2011; Araci, 2019).

LLMs such as BERT (Devlin et al., 2019), FinBERT (Araci, 2019), RoBERTa (Liu et al., 2019), GPT (Brown et al., 2020), OPT (Zhang et al., 2022) and LLaMA (Touvron et al., 2023) have transformed sentiment analysis by moving beyond traditional lexicon-based approaches (Kirtac and Germano, 2024b). These models enable a more nuanced understanding of financial text by considering context, sentiment intensity, and domain-specific language usage (Araci, 2019; Yang et al., 2020; Zhang et al., 2022). As a result, sentiment analysis has become an integral tool for traders, portfolio managers, and researchers in predicting market behavior.

This article provides an overview of financial sentiment measurement methods, the evolution of sentiment analysis in finance, and how LLMs such as BERT-type and GPT-type models calculate sentiment from financial text. While both types of models are designed for natural language processing, they differ in their architectures and methods of understanding text.

BERT-type models, including RoBERTa and FinBERT, employ a bidirectional approach, meaning they analyze words in relation to both preceding and succeeding words in a given sentence. This enables them to capture intricate dependencies within financial text, leading to more precise sentiment classification. RoBERTa enhances BERT's original architecture by eliminating the next-sentence prediction objective and training on longer text sequences, making it particularly effective for structured financial data analysis.

Conversely, GPT-type models, such as OPT and LLaMA 3, adopt an autoregressive framework, predicting sentiment and generating responses based on sequential token probabilities. OPT is optimized for processing large-scale financial datasets with greater computational efficiency, while LLaMA 3 enhances generative capabilities, making it particularly useful for real-time sentiment interpretation. These models are highly effective for analyzing evolving market narratives, summarizing sentiment trends dynamically, and generating predictive insights based on financial text data.

The distinction between these two model types is essential for financial sentiment analysis. While BERT-type models like RoBERTa excel in structured classification tasks, such as categorizing earnings reports and analyst sentiment, GPT-type models like OPT and LLaMA 3 provide advantages in adaptive learning, real-time sentiment generation, and market trend analysis.

## 2. Literature review: measuring financial sentiment

### 2.1 The evolution of sentiment measurement in finance

Financial sentiment has been analyzed using a variety of methodologies, ranging from market-based indicators to survey-based measures and text-based sentiment analysis (Baker and Wurgler, 2006; Brown and Cliff, 2004; Tetlock, 2007; Heston and Sinha, 2017; Manela and Moreira, 2017; Kirtac and Germano, 2024a,b).

Market-based sentiment indicators rely on observable financial metrics, such as trading volume, stock price volatility, IPO activity, and derivatives trading. Baker and Wurgler (2006) constructed a sentiment index that incorporates variables like equity issuance trends, mutual fund flows, and closed-end fund discounts, demonstrating that these indicators predict stock returns. Brown and Cliff (2004) further showed that investor optimism and pessimism influence asset pricing, with excessive sentiment often leading to stock market mispricing. More recent research has extended these measures to include options trading activity and volatility indices, confirming their relevance in assessing market sentiment risk (Jegadeesh and Wu, 2013).

Survey-based sentiment measures provide another approach, relying on investor surveys and behavioral assessments to gauge market expectations. Indices such as the Investor Sentiment Index and American Association of Individual Investors (AAII) Sentiment Survey aggregate self-reported investor outlooks on market conditions (Baker and Wurgler, 2006). While these measures are widely used in behavioral finance, they are often subject to biases, limited sampling, and time lags, making them less effective for real-time market applications (Manela and Moreira, 2017).

Advancements in natural language processing and machine learning have shifted sentiment analysis toward text-based methodologies, allowing researchers to extract investor sentiment from financial news, earnings reports, and corporate disclosures. Tetlock (2007) demonstrated that negative sentiment in financial news articles correlated with subsequent stock price declines, highlighting the impact of media narratives on market movements. Schumaker and Chen (2009) expanded this research by analyzing breaking news headlines, showing that sentiment-derived features could enhance stock return forecasting. These approaches were further refined by Li (2010) and Loughran and McDonald (2011), who developed domain-specific sentiment dictionaries tailored for financial text analysis, improving sentiment classification accuracy.

Researchers have explored how investor sentiment from Twitter, StockTwits, and financial discussion forums impacts market behavior with the rise of social media platforms. Bollen et al. (2011) found that mood states extracted from social media platforms could predict stock market fluctuations, demonstrating the influence of real-time investor discussions. Jegadeesh and Wu (2013) found that analyst reports and corporate disclosures play a crucial role in shaping investor expectations and asset pricing, further validating the role of text-based sentiment measures. Oliveira et al. (2017) confirmed that StockTwits sentiment provides early signals of investor sentiment shifts, offering advantages for short-term trading strategies.

Despite the growing importance of sentiment analysis in financial research, one fundamental challenge remains: the lack of a unified definition of sentiment. Some studies define sentiment as investor optimism or pessimism (Tetlock, 2007; Karoui, 2017), while others view it as a behavioral bias affecting investment decisions (Brown and Cliff, 2004). This lack of consensus has led to inconsistent sentiment indices, where different methodologies often produce conflicting market signals.

Recent studies have integrated multiple sentiment sources to improve predictive accuracy to address these limitations. Heston and Sinha (2017) proposed a multi-source sentiment model incorporating news sentiment, earnings call sentiment, and social media sentiment, demonstrating improved predictive power for stock returns. Kirtac and Germano (2024a) expanded on this by applying LLMs, such as FinBERT and GPT-based architectures, to sentiment classification, showing that LLM-driven sentiment indicators significantly outperform traditional dictionary-based sentiment indices. Their research highlights the superior accuracy of deep learning-based sentiment extraction, particularly in the context of algorithmic trading and quantitative investment strategies.

As LLMs such as BERT, RoBERTa, and OPT continue to evolve, financial sentiment analysis is shifting toward deep contextual learning. These models enable more accurate and dynamic sentiment scores, capturing the nuances of financial language, economic narratives, and market dynamics. The increasing reliance on real-time sentiment insights has made sentiment analysis an indispensable tool for hedge funds, investment banks, and quantitative traders, reinforcing its role in algorithmic trading, portfolio management, and risk assessment.

**2.2 Sentiment constructs and their measurement in prior literature**

Various sentiment constructs have been proposed in financial sentiment research, broadly categorized into general market sentiment and firm-specific sentiment. General market sentiment captures the collective emotions and attitudes of investors, reflecting broader market trends such as optimism, pessimism, and risk appetite (Brown and Cliff, 2004; Baker and Wurgler, 2006). Firm-specific sentiment, on the other hand, focuses on individual companies by analyzing sentiment in corporate earnings reports, analyst recommendations, and news coverage (Tetlock, 2007; Jegadeesh and Wu, 2013).

Lexicon-based sentiment measures rely on predefined dictionaries that classify words as positive, negative, or neutral based on their meaning in financial contexts. The Loughran-McDonald dictionary is one of the most widely used lexicons, specifically designed for financial text analysis (Loughran and McDonald, 2011). Unlike general sentiment lexicons such as Harvard's General Inquirer, the Loughran-McDonald dictionary eliminates misclassified terms in financial contexts, improving sentiment classification accuracy in earnings call transcripts, financial disclosures, and analyst reports (Tetlock, 2007). Despite their advantages, lexicon-based measures face limitations in capturing evolving financial terminology, making them less effective in identifying sentiment shifts in real-time markets. These methods also struggle with contextual negations and financial jargon, leading to potential misclassification errors (Malo et al., 2014). Some studies have attempted to enhance lexicon-based sentiment classification by incorporating sentiment modifiers such as word pair relationships and negation detection algorithms (Schumaker and Chen, 2009; Li, 2010). However, even with these improvements, lexicon-based methods remain constrained by their static nature, prompting researchers to explore machine learning-based sentiment models (Heston and Sinha, 2017).

Machine learning techniques have improved sentiment classification by learning patterns from labeled datasets rather than relying on fixed word lists. Studies have employed models such as support vector machines (SVMs), naïve Bayes, and random forest classifiers to classify financial text into sentiment categories (Heston and Sinha, 2017; Schumaker and Chen, 2009). These models process a wide range of linguistic features, including word embeddings, sentence structure, and source credibility, making them more adaptable than lexicon-based approaches (Jegadeesh and Wu, 2013). Supervised learning approaches have been particularly useful in financial news sentiment analysis, where models are trained on manually labeled sentiment corpora to improve accuracy (Tetlock et al., 2008). However, the dependency on labeled datasets presents challenges, as manual annotation is time-consuming, costly, and prone to human bias (Manela and Moreira, 2017). To address these issues, semi-supervised and weakly supervised machine learning models have been introduced, allowing sentiment classifiers to self-train on partially labeled datasets. Research has found that semi-supervised learning improves sentiment classification accuracy, particularly in environments where sentiment labels are limited (Karoui, 2017).

Deep learning and transformer models have revolutionized sentiment analysis by enabling models to understand financial language in context rather than relying on predefined labels. Transformer models such as BERT, RoBERTa, and FinBERT have been widely used in financial sentiment classification, as they process text bidirectionally, capturing long-range dependencies in financial documents (Araci, 2019; Devlin et al., 2019). FinBERT, a financial domain-specific variant of BERT, has demonstrated superior performance in analyzing earnings call transcripts and market news sentiment, making it a powerful tool for sentiment extraction in quantitative finance (Huang et al., 2023). Unlike lexicon-based methods, deep learning models use word embeddings, allowing for greater contextual awareness in sentiment classification (Heston and Sinha, 2017). Large language models such as OPT and LLaMA have also been explored for sentiment-driven trading strategies, demonstrating their ability to extract sentiment signals from large-scale financial news datasets (Kirtac and Germano, 2024a,b). Their study found that LLM-driven sentiment measures significantly outperformed traditional sentiment indices in forecasting stock price movements, reinforcing the growing role of deep learning in financial sentiment research (Kirtac and Germano, 2024a).

Research has combined lexicon-based sentiment scores with machine learning-derived features to enhance sentiment classification for financial news, earnings reports, and social media sentiment tracking (Tetlock, 2007). Given the strengths and weaknesses of different sentiment measures, successive studies have explored hybrid sentiment models that integrate multiple sources of sentiment data to improve classification accuracy (Heston and Sinha, 2017). Multi-source sentiment integration has gained significant attention in recent literature. Studies have proposed sentiment models that combine news sentiment, social media sentiment, and earnings call sentiment, demonstrating greater predictive power in stock return forecasting. The integration of these sources provides a more robust assessment of market sentiment, allowing for more accurate financial predictions (Heston and Sinha, 2017; Karoui, 2017). Kirtac and Germano (2024a,b) extended this approach, demonstrating that LLM-derived sentiment scores significantly enhance financial forecasting accuracy, reinforcing the importance of hybrid sentiment analysis frameworks in financial modeling.

Financial sentiment analysis continues to evolve with recent advancements focusing on real-time sentiment extraction, multimodal sentiment analysis, and domain-specific LLM enhancements. Research is expanding to include sentiment extraction from earnings calls by analyzing both textual content and vocal tone, further improving the reliability of sentiment assessments. Multimodal sentiment analysis is also expected to gain traction by incorporating financial text, speech, and visual financial data to enhance sentiment classification. Domain-specific fine-tuning of large language models on financial corpora will likely improve sentiment extraction, making sentiment analysis an even more critical tool in algorithmic trading, portfolio management, and risk assessment. The increasing reliance on sentiment insights in financial markets suggests that sentiment analysis will remain a cornerstone of quantitative finance research. Future research is expected to focus on real-time financial sentiment extraction, multimodal sentiment analysis, and domain-specific LLM enhancements, further integrating sentiment insights into algorithmic trading, risk assessment, and behavioral finance models.

## 3. How LLMs measure sentiment from context with a step-by-step example

LLMs have transformed financial sentiment analysis by capturing deep contextual meaning rather than relying on simple word counting or lexicon-based methods. Financial sentiment is often subtle and can be impacted by phrasing, negations, and context. These models analyze sentiment by understanding the relationships between words, making them significantly more accurate in detecting market-relevant sentiment in earnings reports, financial news, and social media.

A financial news headline such as *"Apple reports record revenue despite concerns over supply chain disruptions"* demonstrates this complexity. A traditional lexicon-based approach might classify the sentence as negative due to the words "concerns" and "disruptions." However, a large language model like FinBERT recognizes that "record revenue" is the primary sentiment driver, while "despite concerns" mitigates the negative sentiment. It may classify this as positive sentiment overall. This ability to differentiate between primary sentiment drivers and contextually negated terms makes LLMs superior to traditional sentiment models (Araci, 2019; Kirtac and Germano, 2024a,b).

### 3.1 BERT-type models for sentiment analysis

Bidirectional Encoder Representations from Transformers (BERT) (Devlin et al., 2019) and its financially fine-tuned variant, FinBERT (Araci, 2019), as well as RoBERTa (Liu et al., 2019), process text bidirectionally, meaning they consider both preceding and succeeding words to accurately assess sentiment. FinBERT is specifically trained on financial texts, making it particularly effective in analyzing earnings call transcripts, financial disclosures, and analyst reports. RoBERTa, an optimized version of BERT, enhances performance by training on longer sequences and removing the next-sentence prediction objective, improving sentiment classification in structured financial data. These bidirectional models are particularly useful in financial sentiment analysis, where small contextual changes can significantly impact meaning and interpretation.

An earnings call transcript provides an example of this in practice. The phrase *"the company reported higher-than-expected profits, reversing earlier losses"* differs significantly from *"the company reported lower-than-expected profits, continuing earlier losses."* A traditional sentiment model might struggle to differentiate between these sentences due to their similar structure. A BERT-based model recognizes that in the first example, "higher-than-expected profits" has a strong positive weight, reinforced by "reversing losses," while in the second example, "lower-than-expected profits" has a negative weight, amplified by "continuing losses." As a result, FinBERT correctly classifies the first as positive and the second as negative, demonstrating its contextual understanding.

BERT processes sentiment by first tokenizing and embedding words. Words are converted into 768-dimensional numerical vectors that capture meaning. Unlike traditional models, which read words sequentially, BERT applies positional encodings to maintain word order. Segment embeddings are added to differentiate between multiple sentences when necessary. The model then applies a self-attention mechanism, where each word receives an attention weight based on its importance. In a sentence such as *"stock prices soared after strong earnings,"* the word "soared" receives a high attention weight since it strongly contributes to sentiment, whereas "after" has a low weight.

The CLS token at the start of the sentence represents the entire sentence, capturing contextual meaning from the surrounding text in a 768-dimensional vector. Sentiment classification requires reducing this high-dimensional representation to three sentiment categories: positive, neutral, and negative. A fully connected layer processes the 768-dimensional vector and transforms it into three logits, which are raw scores indicating the model's confidence before probability normalization.

For example, if the output logits before softmax are 2.5 for positive sentiment, 1.2 for neutral, and -0.5 for negative, the softmax function converts these logits into probability values that sum to one. This normalization results in probability values of approximately 0.76 for positive, 0.21 for neutral, and 0.03 for negative. The model assigns the sentiment label based on the highest probability, classifying the text as positive with a confidence level of 76 percent.

A sentiment score of 72 percent positive does not indicate that the sentence contains 72 percent positive words or sentiment intensity but rather reflects the model's confidence in its classification. The model determines that there is a 72 percent probability that the sentiment is positive rather than neutral or negative. This probability-based approach ensures that sentiment classification accounts for contextual nuances rather than just word frequency.

A BERT-based model like FinBERT provides a probability distribution over sentiment categories, a final classification based on the highest probability, and optional raw logits, which researchers can analyze or adjust. BERT does not generate a single numerical sentiment score but instead classifies text by assigning probabilities to sentiment categories. This probabilistic classification method allows for more accurate sentiment analysis in financial applications, where subtle differences in wording can significantly affect interpretation.

## 3.2 GPT-type models for sentiment analysis

GPT-type models, including GPT-4, OPT, and LLaMA 3, employ an autoregressive mechanism, meaning they generate text sequentially, predicting the next token in a sequence based solely on preceding tokens. This fundamental difference sets them apart from BERT-based models such as FinBERT and RoBERTa, which process text bidirectionally, capturing context from both the left and right of a word. The sequential nature of GPT-type models makes them particularly effective for financial sentiment forecasting, text generation, and real-time market analysis, where continuous updates and evolving narratives need to be captured dynamically. Unlike BERT-based models that rely on a CLS token to summarize entire sentences, GPT models dynamically compute sentiment by predicting the next most probable word based on preceding context, making them ideal for analyzing sentiment trends in financial news, earnings calls, and market reports.

GPT models begin the sentiment analysis process by tokenizing text and embedding words into high-dimensional vector spaces. The hidden state dimensions vary depending on the model architecture. Smaller GPT models may use a 768-dimensional representation similar to BERT, but larger models like OPT-6.7B operate with hidden states of 4,096 dimensions, and LLaMA 3 extends even further, surpassing 8,192 dimensions in some configurations. GPT-4, as a state-of-the-art model, utilizes even greater dimensional complexity, enabling it to detect intricate patterns in financial language. These embeddings pass through multiple transformer layers, where self-attention mechanisms adjust each token's representation based on previous words. Unlike bidirectional transformers that allow context to influence interpretation from both sides, GPT models restrict their attention to past tokens, ensuring predictions are based only on prior words without knowledge of future ones.

A financial news headline such as "stock market declines amid fears of recession, but analysts remain optimistic" illustrates how GPT-type models interpret sentiment differently from BERT-based models. BERT-based models, which assess words bidirectionally, might classify this as neutral because of conflicting signals from "declines" and "optimistic." GPT-type models, however, process the sequence autoregressively, first recognizing the negative sentiment in "stock market declines" and "fears of recession." As the model encounters "but analysts remain optimistic," it adjusts sentiment weighting dynamically, incorporating the new information sequentially rather than evaluating the statement as a whole from the outset. This allows GPT-type models to generate interpretative responses such as "despite market declines, optimism from analysts suggests potential recovery." This ability to generate nuanced textual interpretations makes GPT models highly valuable for financial sentiment analysis, particularly in summarizing news reports and identifying sentiment trends across large-scale financial data.

GPT-based sentiment analysis differs from traditional classification approaches because it does not categorize text into fixed sentiment labels. Instead of determining whether a sentence is positive, negative, or neutral outright, GPT-type models assign probability distributions over possible sentiment-expressive words based on context. If a financial document contains both optimistic and pessimistic elements, the model assigns probabilities reflecting the sentiment orientation of each component. For example, in the sentence "the company reported strong earnings but issued a weak forecast," the model assigns high probabilities to words like "strong" and "earnings" within a positive sentiment context, while also recognizing "weak" and "forecast" as indicators of negative sentiment. This allows the model to interpret and adjust its understanding dynamically rather than forcing the entire sentence into a single sentiment category.

Sentiment scores in GPT-based models are computed using a softmax function, which converts raw logits into probability distributions over vocabulary tokens. This differs from BERT-based models, where softmax is applied to classify sentiment into predefined categories such as 76% positive, 21% neutral, and 3% negative. In GPT-based models, softmax is applied to predict the probability of specific words that might convey sentiment, such as "optimistic," "concerned," or "uncertain." If the model predicts "optimistic" as the next word after "but analysts remain," the probability of a positive sentiment classification increases. Conversely, if it predicts "concerned," the sentiment score shifts toward negative. This probabilistic approach enables GPT models to generate richer, more context-aware sentiment assessments, making them particularly useful for financial applications requiring real-time sentiment interpretation rather than rigid categorical classification.

Higher-dimensional representations in GPT-type models contribute to their ability to model financial sentiment dynamically. Unlike BERT-based models like FinBERT and RoBERTa, which use fixed embeddings and a sentence-wide classification approach, GPT models operate in significantly larger vector spaces, allowing them to track sentiment shifts across longer sequences. For instance, GPT-4's extensive hidden states allow it to model evolving financial sentiment across multiple sentences and paragraphs, making it particularly well-suited for long-form market trend analysis and investment decision support. This capability is crucial for financial applications where investor sentiment is shaped by complex, multi-sentence narratives, such as earnings reports, analyst briefings, and regulatory announcements.

GPT-type models also excel at generating human-like explanations of sentiment rather than simply outputting sentiment scores. When processing financial texts, GPT models can produce context-aware responses such as "although the company reported revenue growth, investor sentiment remains cautious due to weak forward guidance." This output differs significantly from traditional sentiment classifiers, which would simply label the text as neutral due to the presence of both positive and negative elements. The ability to provide explanatory sentiment analysis makes GPT-type models invaluable for financial analysts, hedge funds, and algorithmic trading firms that rely on nuanced sentiment insights rather than basic sentiment categorization.

The adaptability of GPT models to various financial contexts further enhances their effectiveness in sentiment analysis. Unlike traditional models that require extensive fine-tuning to adapt to new financial jargon or emerging market trends, GPT-type models can generalize sentiment interpretation across different industries, asset classes, and economic conditions with minimal retraining. Their ability to dynamically process financial text enables them to detect shifts in investor sentiment in response to major events such as Federal Reserve announcements, corporate earnings releases, and geopolitical developments.

Research has consistently shown that GPT-based sentiment models outperform traditional sentiment classifiers in real-time financial sentiment tracking and unstructured market data processing. The flexibility of these models to generate textual explanations rather than merely categorizing sentiment into rigid classes makes them an essential tool for hedge funds, asset managers, and financial analysts seeking to extract meaningful sentiment signals from financial news, earnings transcripts, and analyst reports. Their ability to operate with extensive dimensionality, capture subtle variations in sentiment, and produce human-like sentiment assessments ensures their continued prominence in financial sentiment analysis. These advantages position GPT-type models as indispensable in applications requiring textual generation, dynamic sentiment forecasting, and real-time financial market sentiment interpretation.

### 3.3 Comparing BERT-type and GPT-type models in financial sentiment analysis

BERT-type models such as FinBERT and RoBERTa and GPT-type models such as GPT-4, OPT, and LLaMA are designed for different but complementary tasks in financial sentiment analysis. BERT-type models excel in structured sentiment classification, making them highly effective for tasks that require precise categorization, such as analyzing earnings reports, investor sentiment scores, and financial disclosures. These models utilize a bidirectional attention mechanism, allowing them to process both preceding and succeeding words in a text, enabling more accurate sentiment classification (Devlin et al., 2019). FinBERT, for example, is specifically fine-tuned on financial texts and is particularly effective at extracting sentiment from structured corporate documents, analyst reports, and SEC filings (Araci, 2019). RoBERTa, an improved variant of BERT, enhances sentiment classification by leveraging a more extensive training process, making it well-suited for financial sentiment analysis where fine-grained sentiment distinctions are required (Liu et al., 2019).

In contrast, GPT-type models such as GPT-4, OPT, and LLaMA are better suited for financial text generation, sentiment explanation, and real-time market sentiment interpretation. Unlike BERT-type models, GPT-based architectures follow an autoregressive approach, processing words sequentially rather than considering context from both directions. This design enables them to generate coherent and context-aware financial summaries, making them ideal for tasks such as market sentiment reporting, investor communication, and summarization of breaking financial news (Brown et al., 2020; Zhang et al., 2022; Touvron et al., 2023). For example, a GPT-4 model can dynamically generate sentiment summaries of earnings reports or central bank statements, helping investors quickly assess market sentiment shifts without requiring manual text analysis.

A hedge fund aiming to classify historical earnings call sentiment would likely rely on FinBERT to categorize transcripts as bullish, bearish, or neutral, providing structured sentiment scores for quantitative investment strategies (Kirtac and Germano, 2024a). In contrast, a financial news platform processing real-time market events would benefit from GPT-4's ability to generate sentiment-based narratives, capturing shifts in investor mood and summarizing evolving financial trends dynamically. The generative capabilities of GPT-type models make them particularly valuable for producing human-like textual explanations of sentiment rather than just assigning a fixed sentiment score (Kirtac and Germano, 2024b).

Recent research suggests that integrating both BERT-type and GPT-type models offers the best predictive performance in financial sentiment analysis (Heston and Sinha, 2017; Kirtac and Germano, 2024b). A hybrid approach, where FinBERT or RoBERTa classifies structured financial text while GPT-type models generate textual explanations and market summaries, provides a more comprehensive sentiment analysis framework. This combined methodology enhances sentiment interpretation, allowing for both quantitative scoring of sentiment and the generation of explanatory financial insights. As financial markets become increasingly dependent on real-time sentiment analysis, leveraging the strengths of both model types is expected to improve the accuracy and depth of sentiment-based investment strategies.

## 4. Conclusion

Financial sentiment analysis has evolved into a crucial area of research, with applications ranging from stock market prediction to algorithmic trading. Traditional sentiment measures, such as market-based indicators and survey-based indices, have provided valuable insights into investor behavior. However, advancements in natural language processing and machine learning have enabled more sophisticated text-based sentiment models, improving accuracy and predictive power. The integration of financial news, earnings reports, and social media sentiment into market forecasting models has demonstrated that sentiment-based trading strategies can enhance investment decision-making (Tetlock, 2007; Loughran and McDonald, 2011; Kirtac and Germano, 2024).

Large language models have revolutionized financial sentiment analysis by incorporating contextual understanding and deep learning-based representations of financial text. Unlike traditional lexicon-based approaches, LLMs such as BERT, RoBERTa, FinBERT, GPT, OPT, and LLaMA process sentiment dynamically, capturing nuances in investor sentiment that were previously difficult to quantify. BERT-based models have proven effective in structured sentiment classification, such as analyzing corporate earnings call transcripts and financial disclosures, while GPT-based models excel in generative sentiment analysis and real-time market trend interpretation (Araci, 2019; Devlin et al., 2019; Brown et al., 2020).

The distinction between bidirectional transformer models and autoregressive generative models underscores the importance of selecting the right LLM for financial sentiment analysis. FinBERT and RoBERTa provide high accuracy in sentiment classification tasks, making them suitable for structured financial text, whereas GPT-based models, such as OPT and LLaMA, are better suited for adaptive learning and dynamic sentiment forecasting. Studies have shown that hybrid approaches, which integrate structured sentiment classification with generative modeling, offer superior predictive performance in stock return forecasting and algorithmic trading strategies (Kirtac and Germano, 2024a,b).

As financial markets become increasingly driven by algorithmic decision-making, the role of sentiment analysis in investment strategies continues to expand. Future research is expected to focus on real-time sentiment extraction, multimodal sentiment analysis incorporating speech and visual financial data, and further enhancements in domain-specific LLMs tailored for financial applications. The continuous refinement of financial language models will likely enhance the accuracy of sentiment-driven trading strategies, further solidifying sentiment analysis as an indispensable tool in quantitative finance and behavioral market research.